# Stable Dispersion of Coal Fines during Hydraulic Fracturing Flowback in Coal Seam Gas Reservoirs—An Experimental Study

Faisal Ur Rahman Awan,* Alireza Keshavarz,* Hamed Akhondzadeh, Sarmad Al-Anssari, Ahmed Al-Yaseri, Ataollah Nosrati, Muhammad Ali, and Stefan Iglauer

**ABSTRACT:** In subterranean coal seam gas (CSG) reservoirs, massive amounts of small-sized coal fines are released during the production and development stages, especially during hydraulic fracturing stimulation. These coal fines inevitably cause mechanical pump failure and permeability damage due to aggregation and subsequent pore-throat blockage. This aggregation behavior is thus of key importance in CSG production and needs to be minimized. Consequently, such coal fines dispersions need to be stabilized, which can be achieved by the formulation of improved fracturing fluids. Here, we thus systematically investigated the effectiveness of two additives (ethanol, 0.5 wt % and SDBS, 0.001 and 0.01 wt %) on dispersion stability for a wide range of conditions (pH 6−11; salinity of 0.1−0.6 M NaCl brine). Technically, the coal suspension flowed through a glass bead proppant pack, and fines retention was measured. We found that even trace amounts of sodium dodecyl benzene sulfonate (SDBS) (i.e., 0.001 wt %) drastically improved dispersion stability and reduced fines retention. The retention was further quantified by fractal dimensional analysis, which showed lower values for suspensions containing SDBS. This research advances current CSG applications and thus contributes to improved energy security.

## 1. INTRODUCTION

As the world population increases and human development evolves, the demand for energy also rises. Global gas consumption and demand are currently soaring and will almost be equal to oil consumption by 2040.[1,2] As conventional gas reservoirs are being depleted, unconventional gas reservoirs have formed a pivotal role in meeting global energy demands, especially following their viable commercial development through the introduction of advanced technologies in hydraulic fracturing (HF).[3−5] Coalbed methane (CBM) or coal seam gas (CSG) reservoirs are unconventional reservoirs hosting methane in subterranean coal seams. CSG reservoirs have thus gained popularity in recent decades in Australia, Canada, China, India, and the United States.[6−8] Coal seam gas reservoirs consist of reasonably low-strength rocks, susceptible to failure during drilling (notably directional drilling), hydraulic fracturing, and production.[9] These reservoirs naturally have low permeabilities, and thus, HF stimulation is required in order to develop these resources.[10] Hydraulic fracturing enhances the permeability of coal seams but also reduces the strength and rock mass of a CSG reservoir.[11] CSG production is substantially influenced by coal permeability[12−15] where contributing factors include geomechanical stress regime, gas desorption,[3] clay swelling, mineral dissolution, and precipitation, as well as fines migration.[13,16]

Fines migration has been demonstrated as one of the primary permeability reduction sources in CSG reservoirs.[17,18] The problem with fines migration and subsequent deposition and blockage, including permeability impairment, is non-negligible in CSG.[19] Coal fines, also known as coal dust, coal particles, or pulverized coal, are hydrophobic particles with a wide range of sizes varying from tens of nanometers to tens of micrometers.[8,20,21] The size and amount of migrating coal fines are linked to the well development and operating stage.[8] Coal fines migration initiates with fines generation then causes movement/migration in the reservoir/proppant pack with the flow and culminates in either clogging in a proppant pack or production of the fines at the surface, as schematically illustrated in Figure 1a,b. Coal fines in a hydraulically fractured well tend to cause production issues, block the flow paths in the proppant pack, reduce fracture conductivity, and can prompt production equipment failure.[22] Coal fines generation and migration cause reductions in fracture conductivity and fracture length, thereby impairing the dewatering process.[23] A permeability decline of 35% has been shown to occur when bituminous coal fines were subjected to water flow,[13] while another study observed that conductivity is reduced by 24.4% when only 2% of coal fines flowed into a proppant pack.[23] Therefore, coal fines pose a significant threat to the permeability and production of methane from coal seams where minimizing this phenomenon would be hugely beneficial to CSG extraction.

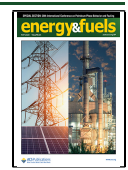

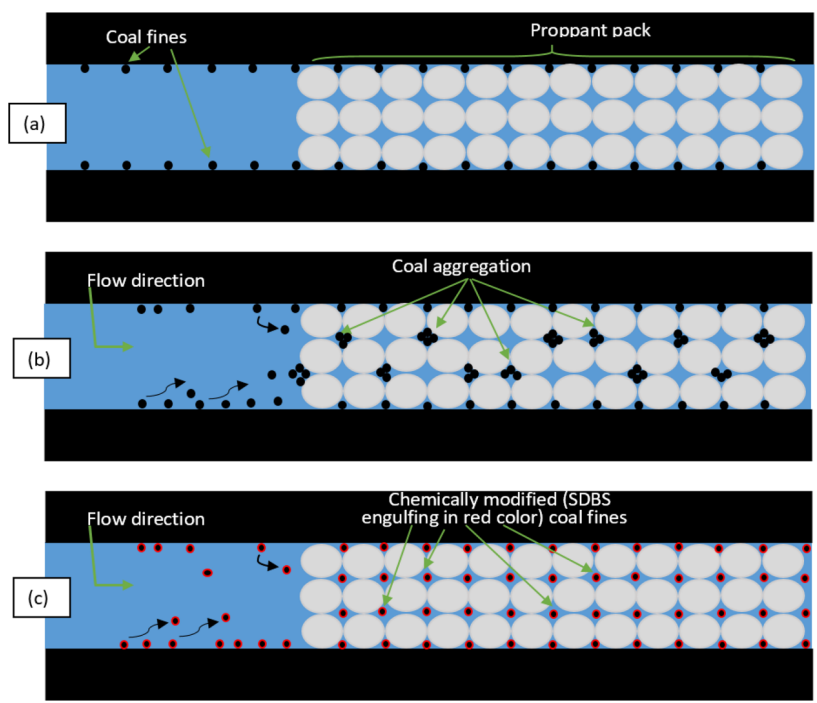

**Figure 1.** Illustration of (a) coal fines generation, (b) coal fines clogging and flow within a proppant pack, and (c) coal fines dispersed using SDBS dispersant.

There are two fundamental approaches to deal with this coal fines challenge: (i) fixating the fines at the source (fines fixation)[24−27] and (ii) managing the released coal by dispersing it in the fracturing fluid for smooth movement of the coal fines through the proppant pack toward the wellbore and production facilities.[28−30] Although fines fixation techniques can reduce the releasing of fines, a significant amount of fines can find its way out of its original place. Consequently, it is crucial to manage released fines. One fundamental way to control the released fines is to chemically modify the fracturing fluid by optimizing the pH and salinity, as well as adding a chemical additive as a dispersant. Accordingly, the dispersants are injected along with the hydraulic fracturing (HF) fluid in order to disperse and produce the released coal fines.[28−30]

The HF suspension contains three essential parts, i.e., water, proppant, and additives. Such additives, constituting less than 1% by weight, comprise gelling agents, pH stabilizers, crosslinkers, corrosion inhibitors, clay stabilizers, well pretreatments, bactericides, and iron sequestrants.[31] Typically, HF fluid applied to coal seams contains specialized dispersing agents to push coal fines through the proppant pack toward the wellbore. Such dispersing agents are known as agglomerate busters.[28]

Characteristically, dispersing agents are classified into three categories, including anionic dispersants,[23] cationic dispersants, and non-ionic dispersants.[23,30] Sodium dodecyl benzene sulfonate (SDBS) has been recently proposed to mix with the HF fluids in CSG reservoirs. As an anionic dispersant, it can (i) alter the wettability of coal surfaces,[32,33] (ii) accelerate the hydrate formation rate in CSG as a kinetic promoter,[34] and (iii) function as a neutralizing and dispersing agent for coal.[35] On the other hand, ethanol has been suggestively used to deagglomerate coal fines dispersion.[25] However, in the studies to date, the synergistic influences of SDBS and ethanol on the dispersion of coal fines were poorly investigated, and the reported data is limited and inconsistent.[23,32,33] Table 1 provides a summary of previous studies using SDBS for controlling coal fines. It is, therefore, essential to systematically study the effects of SDBS and ethanol in the HF fluids in CSG considering all potential scenarios.

In this work, we thus conducted a comprehensive experimental study to develop an optimized recipe for the HF fluids in CSG reservoirs by optimizing salinity, pH, added SDBS, and ethanol to the base fluids in order to improve the coal fines dispersion and smooth movement of the suspension through the proppant pack. The results are consistent with one another and have excellent repeatability. This paper introduces an effective solution to disperse coal fines by adding an anionic dispersant of specific concentration to the HF fluid and gives the first insight into the phenomenon of coal fines dispersion via SDBS priming where an illustration of this phenomenon can be seen in Figure 1.

## 2. EXPERIMENTAL METHODOLOGY

In order to identify the optimal formulations for coal fines dispersion, a series of experiments were systematically conducted to study the effect of salt, ethanol, SDBS concentrations, and pH on the dispersion of coal fines in the liquid phase. To achieve this, various tests (proppant flow test, zeta potential measurements, particle analyses, and microscopy imaging) were conducted. Coal water suspensions of known pH to ensure the similarity of conditions, i.e., 8.5 ± 0.1 with various salinities were formulated as discussed below, which were subsequently analyzed using a particle counter sizer. The pH was selected based upon the reported properties of intact coal seam formation waters.[38] Each suspension was analyzed after permeating through the glass bead proppant column (as can be seen in Figure 2) via conducting zeta potential measurements and particle size analysis followed by microscopy imaging. To further augment the findings, we also analyzed the raw coal, SDBS, and modified coal fines (treated with SDBS) with a Perkin Elmer FTIR spectrometer. Exclusively, the materials and methods applied are discussed in detail below.

**2.1. Materials.** A highly volatile bituminous coal[39] received from Morgantown, West Virginia, USA was used in this study. This coal sample has a relative density of 1.31, a vitrinite reflectance of 0.91, and maceral and mineral composition of 97.4 and 2.6 vol %, respectively; more details can be found in Table 2. The received bulk coal was crushed with a mortar and pestle where the coal fines were then sieved on an electric sieve shaker for 20 min. The particle size fraction, 20−38 μm, of raw, dry coal fines was used in all tests, i.e., for the zeta potential analysis, particle size analysis, proppant flow testing, and microscopy imaging.

Table 1. Summary of Experiments Employing SDBS on Coal Fines

| s # | coal properties | base fluid | surfactants | studies | results |
|---|---|---|---|---|---|
| 1[36] | carbon black<br>$d_{av} \approx 2.33\ \mu m$<br>$\varsigma = -5.7$ mV<br>$C = 0.1$ wt % | saline water based in NaCl (0 to 0.2 M) | SDBS [$C = 0.0034$ to $0.0445$ wt %] | interaction of SDBS on $\varsigma$ and adsorption isotherms | 0.01 wt % SDBS gives an optimum change in $\varsigma$ in all NaCl salinities tested from 0 to 0.2 M |
| 2[23] | coal fines, $d_{av} = 45$ and $150\ \mu m$<br>$\varsigma = -20.5$ mV<br>$C = 1$ wt % | saline water based in KCl (0 and 0.27 M) | FSJ-01, FSJ-02, SDBS, MF, NNO [$C = 0.01$ wt %]<br>Tween-80, Span-80 [$C = 0.1$ wt %] | interaction of 7 surfactants with the coal fines in proppant pack | 0.01 wt % FSJ-02 gives an optimum change in $\varsigma$ with wettability reversal, the recovery rate improved by 31.5%, and conductivity improved by 13.3% |
| 3[30] | coal fines<br>$d_{av} = 45$ and $150\ \mu m$<br>$\varsigma = -30.5$ mV<br>$C = 1$ wt % | saline water based in KCl (0 and 0.27 M) | FSJ-01, FSJ-02, SDBS, MF, NNO [$C = 0.01$ wt %]<br>Tween-80, Span-80 [$C = 0.1$ wt %] | interaction of 7 surfactants with the coal fines in proppant pack | 0.01 wt % FSJ-02 gives an optimum change in $\varsigma$ (approximately $-60$ mV) with wettability reversal, coal fine recovery rate and pack conductivity improved. |
| 4[32] | sub-bituminous Collie coals<br>$d_{av} \approx 30-40\ \mu m$<br>$\varsigma \approx -10$ mV<br>$C = 0.05$ wt % | DI water | SDBS [$C = 0.05$ to $1.0$ wt %] | interaction of SDBS on $\varsigma$, wettability, adsorption | 0.2 to 0.4 wt % SDBS gives optimum change in $\varsigma$ and surface tension. |
| 5[37] | bituminous coal<br>$d_{av} = 20-38\ \mu m$<br>$\varsigma \approx -30$ mV<br>$C = 0.05$ wt % | saline water based in NaCl (0, 0.1, and 0.6 M) | SDBS [$C = 0.01$ wt %] | interaction of SDBS on $\varsigma$ and retention in packs | 0.01 wt % SDBS gives optimum change in $\varsigma$ in Lower NaCl salinity tested i.e. 0.1 M |

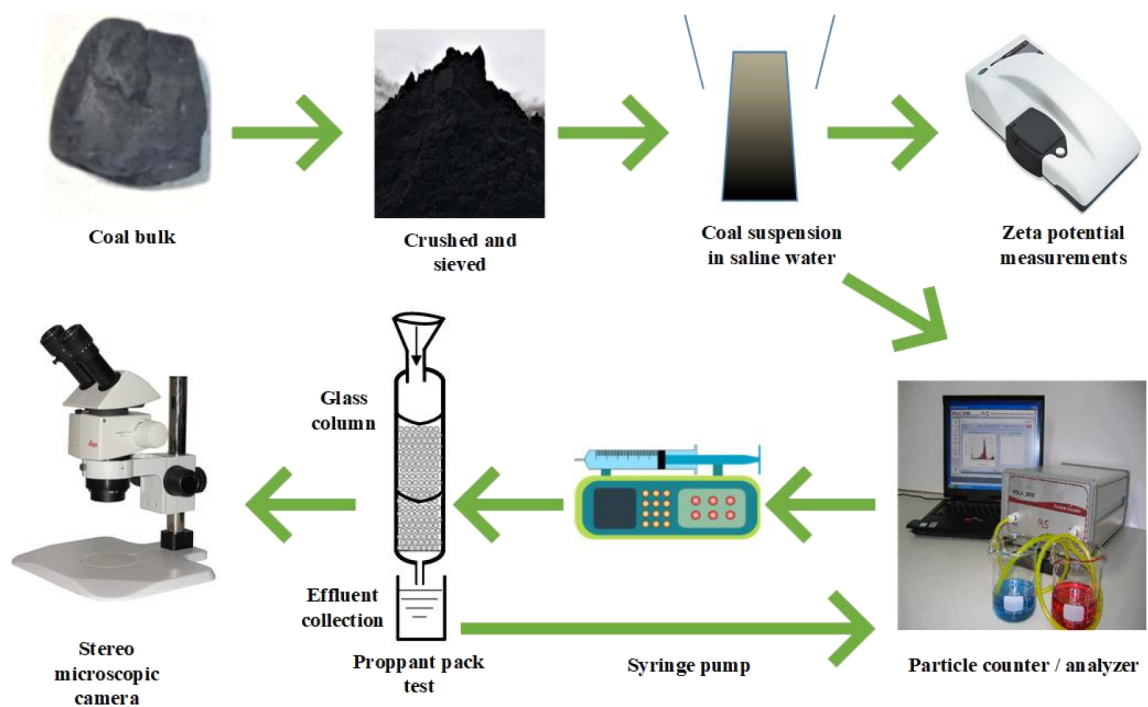

**Figure 2.** Schematic diagram of glass bead proppant pack flow test, zeta potential, particle counter, and microscopy camera.

**Table 2. Properties of the Coal Sample (as Tested by Bureau Veritas - Minerals Pty Ltd)**

| parameter | standard | value |
|---|---|---|
| ash (%, db[a]) | BS 1016.104.4 (1991) | 4.80 |
| density (g/cm$^3$) | AS 1038.21.1.1 | 1.31 |
| volatile matter (%, daf[a]) | BS 1016.104.3 (1991) | 5.042 |
| C (%, daf[a]) | BS 1016.6 | 78.50 |
| H (%, daf[a]) | BS 1016.6 | 5.37 |
| $R_v$, max[a] (%) | AS2486.3 (2000); 546NM; Oil RI 1.518; Standards 0.29−1.71% | 0.91 |
| vitrinite (vol % mmf[a]) | AS2856.1 (2000), AS2856.2 (1998), AS2856.3 (2000), international ISO7404 and ICCP guidelines | 81.01 |
| inertinite (vol % mmf[a]) | AS2856.1 (2000), AS2856.2 (1998), AS2856.3 (2000), international ISO7404 and ICCP guidelines | 12.73 |
| liptinite (vol % mmf[a]) | AS2856.1 (2000), AS2856.2 (1998), AS2856.3 (2000), international ISO7404 and ICCP guidelines | 6.26 |

[a]db: On a dry basis; daf: dried ash-free; $R_v$, max: maximum vitrinite reflectance; mmf: mineral matter-free.

Deionized (DI) water produced by the Millipore Direct-Q 3UV that supplied ultrapure (type 1) water was used as the base fluid in all of the experiments. A diluted aqueous solution of NaOH (0.01 M) (Rowe Scientific Pvt. Ltd.) was used for pH adjustment of the coal formulations as required. The sodium chloride (NaCl) analytical reagent (Rowe Scientific Pvt. Ltd.) was used for adjusting the ionic strength of the base fluid. Sodium dodecyl benzene sulfonate (SDBS) ($C_{18}H_{29}NaO_3S$) of technical grade (Sigma-Aldrich Pty) and ethanol absolute analytical reagent ($C_2H_5OH$) (Rowe Scientific Pvt. Ltd.) were used as additives.

Ethanol has been suggestively used to deagglomerate coal fines dispersion.[25] Ethanol concentration was established on a Mastersizer 3000 by adding 0−2 wt % ethanol in a beaker of 600 mL of water suspension. Resultantly, 0.5 wt % ethanol yielded a minimum averaged size of coal particles of ∼19.07 μm, as can be seen in Figure 3. This weight of 0.5 wt % ethanol was used to further investigate the effect of ethanol on coal fines aggregation behavior.

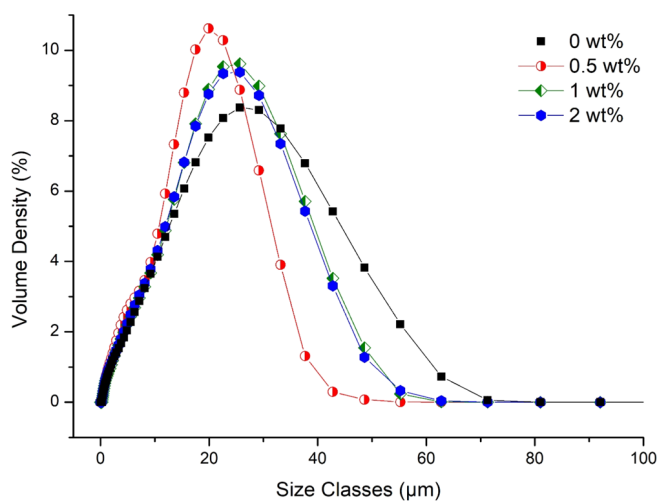

**Figure 3.** Effect of ethanol concentration (in wt %) on coal fine particle size.

SDBS, an anionic dispersant, has been used by several researchers to stabilize coal fines dispersion. The zeta potential of coal in the presence of SDBS has been investigated and reported as (i) ζ of −28 mV in an aqueous 2 wt % KCl brine with 1 wt % coal fines and 0.01 wt % SDBS,[23] (ii) −55 mV in a non-saline suspension of 0.05 wt % coal fines in 0.05 wt % SDBS,[32] and (iii) −32 mV in aqueous 0.1 M NaCl brine with 0.1 wt % coal fines and 0.41 wt % SDBS.[33] Variable coal fines compositions (ranging from 0.05 to 1.00 wt %) in the literature have resulted in diversified zeta potentials from −28 to −70 mV when SDBS is used. An optimum concentration of 0.05 wt % SDBS has been shown to optimize the stability of the nano-formulation (containing copper and alumina NPs) and also to reduce the particle size of nanoparticles over a wide pH range.[40] Accordingly, in this research, we used the lowest tested SDBS concentration in the literature (0.01 wt %),[30,37] as can be seen in Table 1, and upon observing good results, reduced it further to 0.001 wt %.

**2.2. Particle Characterization Using Zeta Potential Measurements.** Zeta potential is a key indication for colloidal stability[41] since it controls the particle's ability to resist the collision, coalescence, aggregation, and the subsequent sedimentation.[42,43] Characteristically, dispersions with a zeta potential below ±5 mV are unstable, while dispersions with zeta potentials ranging from ±10 to ±30 mV are incipient stable.[44] Furthermore, dispersions with zeta potentials ranging from ±30 to ±40 mV are moderately stable, and the ones with zeta potentials ranging from ±40 to ±60 mV are considerably stable dispersions.[44] Eventually, dispersions with zeta potentials higher than ±60 mV are significantly stable. Thus, typically, zeta potentials of more than ±40 mV result in stable suspensions.[45] Note that colloidal stability is affected by pH levels, coal fines type, the concentration of coal fines, base fluid, ionic strength, and any additives used.[25]

**Table 3. Average Properties of the Glass Bead Proppant Pack**

| parameter | value |
| --- | --- |
| packing density | 1524 kg/m$^3$ |
| porosity | 40.6 ± 2% |
| pore Volume | 8.1 ± 0.4 mL |

Coal suspensions were formulated by first adding coal fines into a beaker followed by additives (either SDBS or ethanol or both), as shown in Table 4, and finally, brine with different ionic strengths. All components of the suspension were then weighed using a high-accuracy balance (HR-250 AZ with Super Hybrid Sensor; accuracy, ±0.0001 g). Subsequently, to obtain a stable pH value, the formulated coal water suspension (CWS) was stirred with a magnetic stirrer (Cole-Parmer) at 600 rpm at ambient temperature 23 ± 1 °C for 5 min at each pH level. A Starter 3100 pH meter (Ohaus Corporation) was used to measure and monitor the pH of the prepared coal suspensions where the instrument was calibrated daily using three-point calibration (95−98% accuracy). The pH was adjusted on the alkaline side by gradually adding aqueous 0.01 M NaOH. Then, the zeta potential was measured using a Zetasizer Nano series Nano ZS version model ZEN3600 (Malvern Instruments Ltd.) for each salinity at least thrice. Here, the measurement cell of the zeta sizer was filled using a capillary tube until reaching the desired level. Brines based on NaCl with 0.001 M, 0.01 M (very low saline), 0.1 M (low saline), 0.3 M (moderate saline) and 0.6 M (standard saline) were tested for the pH range of 6 to 11.[38]

**Table 4. Experimental Matrix for Zeta Potential Tests, Proppant Pack Test, and Particle Size Distribution**

| suspension name | DI water | NaCl | coal fines (wt %) | ethanol | SDBS |
| --- | --- | --- | --- | --- | --- |
| S-1 | 99.3656 | 0.5844 | 0.05 | | |
| S-2 | 98.8656 | 0.5844 | 0.05 | 0.5 | |
| S-3 | 98.8646 | 0.5844 | 0.05 | 0.5 | 0.001 |
| S-4 | 99.3646 | 0.5844 | 0.05 | | 0.001 |
| S-5 | 99.3556 | 0.5844 | 0.05 | | 0.010 |
| S-6 | 98.1968 | 1.7532 | 0.05 | | |
| S-7 | 97.6968 | 1.7532 | 0.05 | 0.5 | |
| S-8 | 97.6958 | 1.7532 | 0.05 | 0.5 | 0.001 |
| S-9 | 98.1958 | 1.7532 | 0.05 | | 0.001 |
| S-10 | 98.1868 | 1.7532 | 0.05 | | 0.010 |
| S-11 | 96.4436 | 3.5064 | 0.05 | | |
| S-12 | 95.9436 | 3.5064 | 0.05 | 0.5 | |
| S-13 | 95.9426 | 3.5064 | 0.05 | 0.5 | 0.001 |
| S-14 | 96.4426 | 3.5064 | 0.05 | | 0.001 |
| S-15 | 96.4336 | 3.5064 | 0.05 | | 0.010 |

**2.3. Proppant Flow Tests and Microscopy Imaging.** To physically simulate the proppant pack flow, a glass column (400 mm long, 20 mm diameter, and a 19/26 bore, with a thimble mesh of 150 μm at the base) was used. The spherical glass beads (minimum roundness, 65%) were sieved (size fraction, 425−600 μm) and then packed in a glass column. Ultralight and high-strength hollow borosilicate glass particles (Potters Industries Pty. Ltd.) were used as proppant particles, 30 g occupying a volume of 19.65 cm$^3$, as shown in Table 3. Coal fines suspensions containing 0.1 wt % coal were introduced into the pack where coal retention was then analyzed. All suspensions were injected at a constant pH of 8.5 and under ambient conditions. Suspensions before and after glass bead pack permeation were analyzed with the particle counter analyzer for particle size distributions, number of particles, and volume of particles. The retention factor, dimensionless retention, for measuring the amount of CWS that was retained within the proppant pack is

$$\text{retention factor} = 1 - \left[\frac{C_{\text{eff}}}{C_{\text{o}}}\right]$$

where $C_{\text{eff}}$ and $C_{\text{o}}$ are the concentration volume of coal fines in the effluent and influent, respectively.

For microscopy imaging of the proppant after flooding, to visualize the coal fines retention in the proppant pack, a Leica M80 routine stereo microscope was employed with a visibility window dimension of 792 × 594 μm. The CWSs (with and without additives) of 0.1, 0.3, and 0.6 M NaCl ionic strength were introduced and captured in the proppant pack. The images qualitatively represent the impact of additives to pass through the proppant pack.

The images were analyzed and processed with AVIZO by applying filters and Image-J software to determine particle sizes, areas occupied by coal fines, and thus particle agglomeration in the proppant pack. The quantitative results were obtained after filtering the images with Image-J software and segmenting the images into binary files (two-color: black and white), each with a resolution size of 792 × 594 units (1024 × 768 pixels), 8-bit (inverting LUT), 768 K images. The fractal box-count method was employed with box sizes of 2, 3, 4, 6, 8, 12, 16, 32, and 64. Note that the slope (which represents the fractal dimension $D$) is obtained by plotting log (box size) vs. log (count). The fractal dimension ($D$) quantifies the degree of aggregation of coal fines suspension.[46,47] Resultantly, larger aggregates lead to higher $D$ values.

**2.4. Selection of Test Conditions.** In achieving the objective of coal fines dispersion, various suspensions were employed. These suspensions have been formulated in various compositions as (a) base fluid (0.1 M (low salinity),[48,49] 0.3 M, and 0.6 M NaCl salinity), (b) additives used (ethanol, 0.5 wt % and SDBS, 0.01 or 0.001 wt %), and (c) pH range (6−11). The selection procedure of SDBS can be found in Table 1 and ethanol in Figure 3. Thus, the set of experiments can be seen in Table 4.

## 3. RESULTS AND DISCUSSION

Coal fines are hydrophobic and tend to agglomerate with each other where this property is known as hydrophobic flocculation.[50] In CSG reservoirs, coal fines can thus plug the cleat system and drastically reduce the permeability of the reservoir.[51] Coal fines exhibit a negative zeta potential in a reservoir environment (pH range 6−10) where dispersion stability is enhanced with an increase in pH.[23,25,52] Zeta potentials of coal suspensions in DI water were reported as −43.34 mV for bituminous[25] and −20.5 mV for anthracite;[23] however, the pH and salinity have not been studied together systematically. Furthermore, fines suspension with DI water, 0.268 M KCl, and 0.2 M NaCl was reported, which, however, is only circa half of average seawater salinity,[23,30] while hydraulic fracturing fluid is usually saline (nearly 0.6 M NaCl or 0.44 M KCl), and the pH is usually moderately alkaline (8.5) to be chemically compatible with the formation water.[25,38,51] Note that seawater salinity is 32.4 g of KCl/L of water (0.435 M) or 35.0 g of NaCl/L of water (0.599 M).

In contemplation of the understanding of the ionic strength effect, additive concentration (SDBS and ethanol), and coal suspension flow within proppant packs, we systematically examined surface properties of the coal fines. The results lead to a comprehensive understanding of the effects of SDBS dispersant/additive on coal fines dispersion as a function of ionic strength, concentration, and dispersion stability through the proppant pack. This information can be used to optimize CSG fracturing fluid to resolve the issue of fines aggregation.

### 3.1. Effect of Salinity on Coal Fines Dispersion Behavior.
We found a linear trend of ionic strength on the coal fines dispersion where zeta potential ($\zeta$) becomes more negatively charged with an increase in pH and a decrease in ionic strength. However, at pH levels from 9 to 11, the $\zeta$ overlaps or increases infinitesimally, as can be seen in Figure 4, because the electric double layer (EDL) formed at the coal surface expands where the overlap of EDL's like-charge surfaces increases the EDL repulsion energy, which may cause detachment of fines from the surface.[53]

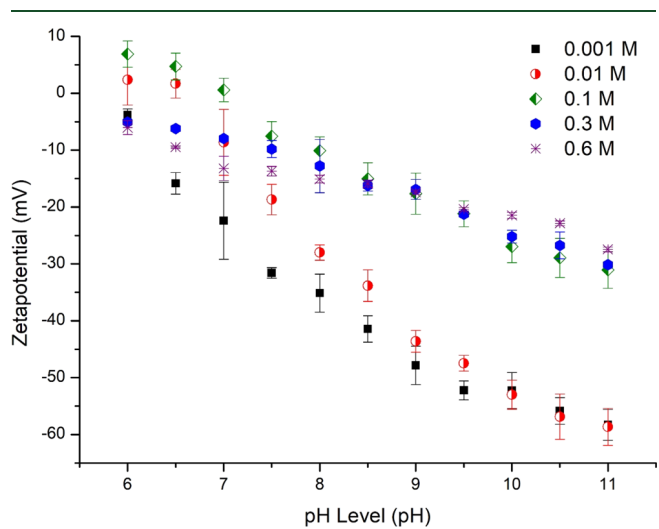

**Figure 4.** Zeta potential–pH relationship of coal fines as a function of salinity.

The zeta potential of the coal fines measured at pH level 8.5 was constant (−15 mV) at this salinity range 0.1–0.6 M NaCl. However, as ionic strength decreases, the zeta potential increases at a particular pH, as can be seen in Figure 4. This indicates that coal fines are hydrophobic ($\zeta$ values within the range of −10 to +10)[50] and tend to agglomerate at higher ionic strengths, especially at standard salinity (i.e., 0.6 M NaCl). Ionic strengths of 0.1, 0.3, and 0.6 M in the range of 6–8 pH were thus highly unstable ($\zeta$ = +10 to −10 mV). Aggregation of coal fines is prone to occur in this unstable suspension due to the ion exchange of Na$^+$ and lower impact of OH$^-$ to counter the effect. Shi et al. (2018) have reported this phenomenon for saline suspensions when the surface of coarser coal fines is attached to finer coal fines, and thus, the suspension stability is reduced.[25] According to the extended DLVO (Derjaguin−Landau−Verwey−Overbeek) theory, when salt (NaCl) is added to water (H$_2$O), the cations (H$^+$ and Na$^+$) can screen the repulsive charge of the EDL and promote coal fines agglomeration (especially when Cl$^-$ ionizes H$^+$),[25] leading to a thin EDL and hence less repulsion between likewise charged particles. Upon adding NaOH, OH$^-$ is hydrolyzed by Cl$^-$, which reduces suspension viscosity[25] and enhances stability, Figure 4. Therefore, in CSG reservoirs, standard salinity will result in greater agglomeration of coal fines and thus permeability reduction. We also conclude that hydraulic fracturing fluid injection of very low ionic strengths (0.01 and 0.01 M) is not feasible in the field: first, because it is not economical (due to excessive filtration); second, because it causes more coal fines to be generated; and third, as it is not geochemically compatible with CSG formation waters. Thus, we focused on low (0.1 M), medium (0.3 M), and standard salinity (0.6 M) at a pH level of 8.5 in the subsequent proppant pack experiments, as these conditions are more prevalent in subterranean coal seams.

### 3.2. Impact of Additives on Dispersion Stability.
The 0.1 M NaCl brine and 0.5 wt % ethanol suspension (S-2) did not significantly improve colloidal stability when compared with coal suspension with no additives (S-1). However, suspension S-3, a CWS, is formulated with 0.5 wt % ethanol and 0.001 wt % SDBS, resulting in enhanced dispersion stability. Contrastingly, as shown below, the $\zeta$ values of ethanol cases are lesser when compared to SDBS only cases. Although researchers (e.g., Shi et al. (2018)) have used ethanol to deagglomerate coal fines dispersions,[25] our results show that ethanol does not improve dispersion stability. We hypothesize that this is because of ethanol's non-electrolytic nature where it does not ionize in aqueous suspensions.

SDBS, however, provides a substantially improved dispersion stability for all studied pH ranges via supercharging the dispersed particles. Low ionic strength (0.1 M) with low SDBS concentration (0.001 wt %) resulted in excellent colloidal stability (absolute $\xi$ potential, >50 mV) over a wide pH range of 6−11. Note, however, that the effect of alkalinity on zeta potential is neutralized when SDBS is used. This is because SDBS is adsorbed on the coal fines surfaces to the maximum adsorption capacity and forms a thick EDL layer where we hypothesize that there is no more space on the coal fines surface available to adsorb more OH$^-$ charges. It was also observed that SDBS concentrations of 0.001 and 0.01 wt % had almost the same effect on the zeta potential−pH relationship, as displayed in Figure 5. We conclude that even a trace amount of SDBS can yield excellent stability of coal fines in CWS.

The negative charges on the surface of the coal particles can attract a dispersant's positive charge; however, the outer coal fines layer is chemically modified by SDBS, which causes an increase in negativity of the surface, resulting in a supercharged coal fines surface, and hence, the repulsion between coal fines increases. Mechanistically, due to the similar charges between coal fines and the SDBS head group, the tail group of SDBS attaches to the coal fines' surface, resulting in higher negatively charged coal fines (as visualized in Figure 6). Such higher surface charges enhance the repulsive forces between the adjusted fines in the dispersion.

Low absolute zeta potentials (±10 mV) on the coal surface indicate a neutrally charged surface; thus, SDBS can adsorb on the coal surface. Thus, at low SDBS concentration, negatively charged dodecyl benzene sulfonate is absorbed on the relatively neutral surface of the coal fine, and consequently, the surface of coal is negatively charged. At higher SDBS concentration, the anionic groups in the adsorbed layer increase the absolute zeta potential, resulting in increased repulsive forces between the particles and an increase in EDL thickness. Furthermore, at a constant SDBS concentration, when the pH is increased above 10, $Na^+$ from NaOH adsorbs on an adsorbed layer of coal coated with SDBS reducing the net surface charge, thus slightly reducing the absolute zeta potential, as seen in Figure 5. There is a maximum adsorption (coating) capacity at the surface of each coal fine. Thus, SDBS partially ionizes in water and gives anionic species, whereas coal has an affinity for anionic groups.[40]

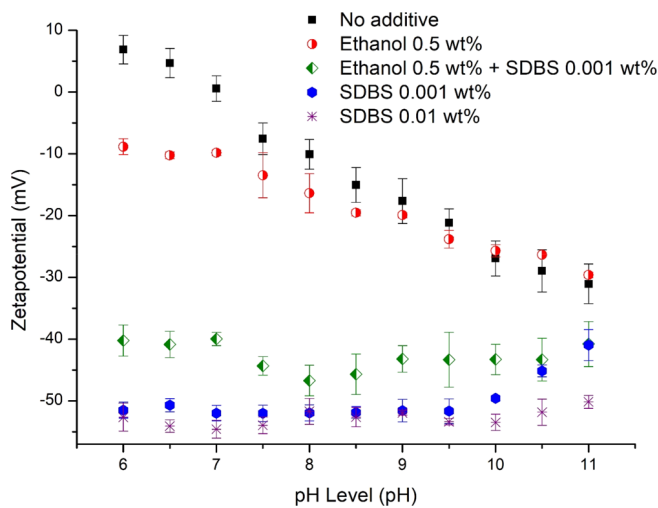

Figure 5. Zeta potential−pH relationship of coal fines for various ethanol−SDBS formulations at 0.1 M NaCl concentration.

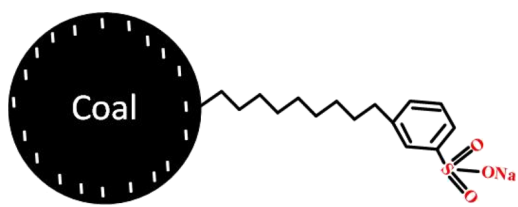

Figure 6. Schematic of SDBS attachment on a coal fines' surface.

To analyze this adsorption effect, FTIR analysis was performed on raw coal fines, SDBS particles, and modified coal fines. The results show that the raw bituminous coal fines contained carbonyl (1700 $cm^{-1}$) and hydroxyl (3100 $cm^{-1}$) surface groups, as displayed in Figure 7. The modified coal fines showed peaks at wavelengths of 1182, 1120, 1032, and 982 $cm^{-1}$ indicating the presence of strong stretching S═O sulfonate groups, stretching of S═O sulfone groups, stretching of S═O sulfoxide groups, and alkene C═C bending groups, respectively, as shown in Figure 7. The FTIR spectra show that all groups present at the SDBS surface are also present on the modified coal fines surface. We conclude that SDBS has modified the coal fines surface. This indicates that modified coal surfaces have a higher negative charge and stronger interparticle repulsive forces.

**3.3. Impact of Salinity on SDBS Performance.** Another set of experiments was carried out to determine the stability of chemically modified coal fines (CMCF), i.e., coal fines treated with SDBS, as a function of ionic strength and pH. The results

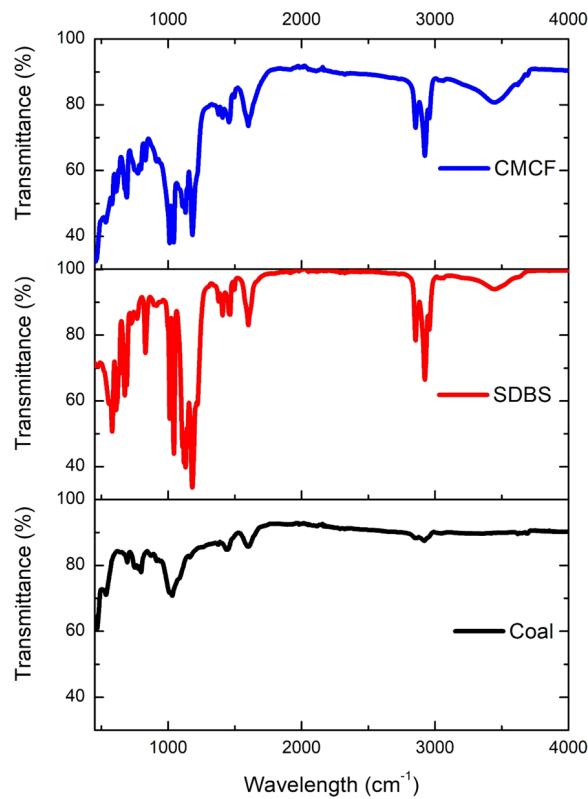

Figure 7. FTIR spectra of CMCF, SDBS powder, and Coal fines.

clearly show that increased ionic strength decreased the zeta potential of CMCF, as can be seen in Figure 8. Thus, CMCF

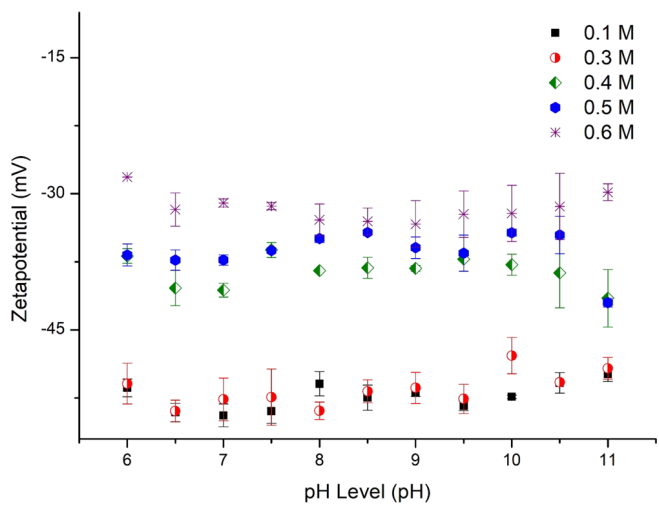

Figure 8. Effect of 0.001 wt % SDBS at various ionic strengths.

has excellent dispersion stability (i.e., ζ potential of −52 mV) at ionic strengths up to 0.3 M. However, increasing the ionic strength to above 0.3 M resulted in lower stability (zeta potentials of −40, −36, and −34 mV at 0.4, 0.5, and 0.6 M, respectively), over a pH range of 6−11, as shown in Figure 8.

At higher salinities (0.4−0.6 M NaCl), the $Na^+$ ions reduced the repulsive forces (which were enhanced by the adsorbed dodecyl benzene sulfonate); hence, we observed lower zeta potentials, as shown in Figure 8. We also observed that zeta potential of SDBS in CWS remains constant in a wide pH range, i.e., zeta potential becomes independent of pH.

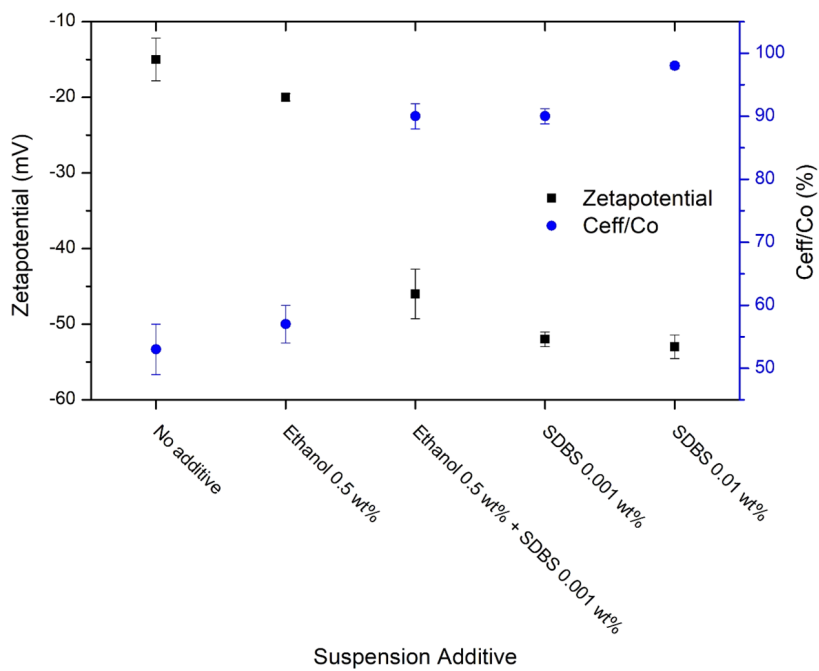

**Figure 9.** Zeta potential−$C_{eff}/C_o$ relationship for various formulations in 0.1 NaCl brine.

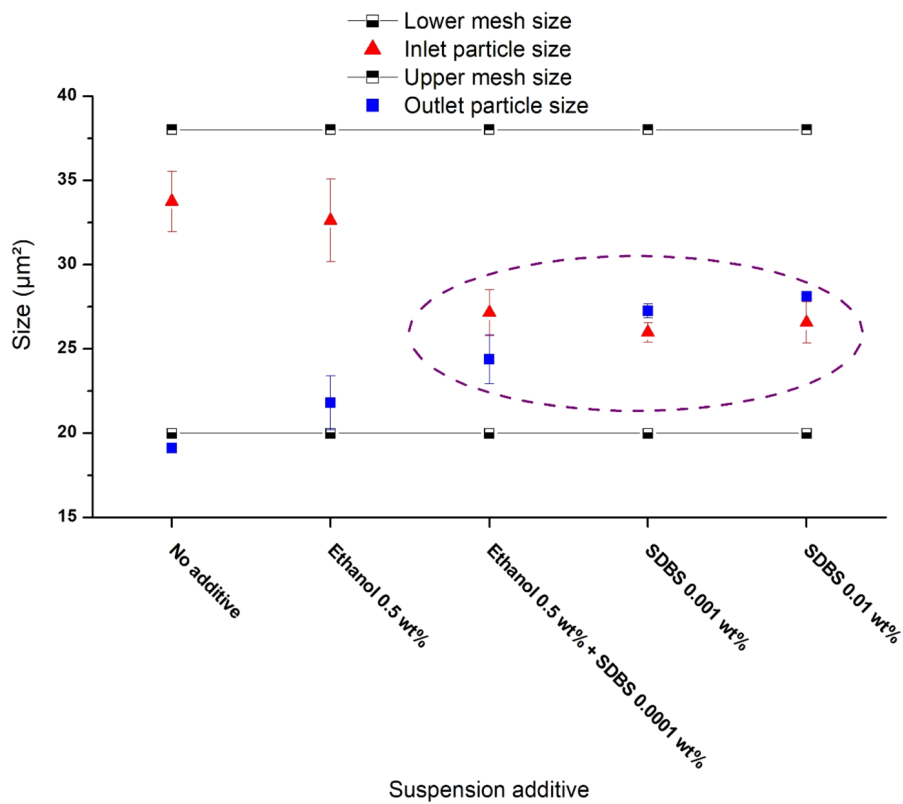

**Figure 10.** Particle analysis at the inlet and outlet of the proppant pack for the various coal water suspension. The purple dashed circle shows that SDBS-treated cases results in an almost similar average particle size.

However, ionic strength has a significant effect on SDBS performance. Thus, at 0.1−0.3 M NaCl ionic strengths, the absolute zeta potential was very high (i.e., >50 mV) for almost the whole pH range tested, while 0.3 M salinity was the highest salinity, which gave excellent results; this salinity is also feasible for use in CSG reservoirs.[54]

**3.4. Dispersion of Coal Water Suspension Using Additives within Proppant Pack.** The performance of the various HF fluids was tested in a synthetic proppant pack before and after modifying the coal fines. A particle analyzer was used, as discussed in section 2.3, to measure the retention factor of the particles after they passed through the glass bead proppant pack. As mentioned earlier, the higher the absolute zeta potential, the better the dispersion stability; hence, we hypothesized that the retention factor would be lower. The zeta potential and the $C_{eff}/C_o$ of coal fines for suspensions of 0.1 M NaCl are plotted in Figure 9.

A higher zeta potential led to a higher $C_{eff}/C_o$ (i.e., lower retention factor), as shown in Figure 9. The coal only and 0.5 wt % ethanol showed a zeta potential of about −15 and −20 mV, respectively, corresponding to $C_{eff}/C_o$ of 53 and 57. However, in cases when SDBS was added, higher absolute zeta potentials (>46 mV) and higher $C_{eff}/C_o$ (>90%) resulted. However, there was not much difference between 0.001 wt % SDBS and 0.01 wt % SDBS.

Another analysis of the above results was conducted on a particle counter sizer for 0.1 M NaCl ionic strength. The results showed that S-1 (CWS containing no additive) possesses the largest average particle size at the inlet (34 μm) and the lowest at the outlet (18 μm). Note that 18 μm is even smaller than the lowest mesh size (i.e., 20 μm), as can be seen in Figure 10. This shows that even dried coal fines tend to partially agglomerate, consistent with previous research.[55] However, the average size of the coal fines at the inlet and outlet is approximately the same whenever SDBS is used, which clearly explains and validates the results obtained from zeta measurement and $C_{eff}/C_o$ that about 90% of coal fines passed through the pack, as seen in Figure 9.

SDBS-treated coal fines dispersions are stable (zeta potential more than −34 mV) not only in suspension but also when passed through the proppant pack. This impedes the aggregation of coal fines and thus helps to transport coal fines through a proppant pack. Dispersion reduces the coal fines' average particle size, which makes it easier to be suspended (within the aqueous environment) and flow back during post hydraulic fracturing operation and thereby increases the productivity and conductivity of the CSG reservoir.

**3.5. Microscopy Imaging.** Images of the proppant pack after CWS passed through the proppant pack were taken through a stereomicroscope and compared qualitatively as well as quantitatively.

Two groups (coal without any additive and coal with 0.5 wt % ethanol) qualitatively showed higher agglomeration of coal fines, while SDBS-added CWS showed well-dispersed coal fines with minimum agglomeration even at higher ionic strength (0.6 M NaCl brine), as can be seen in Figure 11.

Quantitative analysis of the particle size, area, and aggregation reveals that the chemically modified coal fines (CMCF) have minimum particle size, minimum area, the least volume, and negligible aggregation. For all salinities tested (0.1, 0.3, and 0.6 M NaCl), the larger coal fines agglomerated into a broader range of larger agglomerates, while finer coal fines formed smaller agglomerates. For example, at 0.6 M salinity, S-11, S-12, and S-13 formed larger aggregates and with a mean size of 16.54−19.56 μm, while in the case of S-14 and S-15, the mean size was merely 3.51−4.63 μm. SDBS-containing suspensions yielded a lower average diameter of fines, e.g., 3.5 μm for case S-15 and the least area of coal fines, e.g., 2663.6 μm$^2$ in case S-4, as can be seen in Figure 12. The total particle count (more particles refer to good dispersion) and area analysis (lower areas) also imply that suspensions containing SDBS outperform all ionic strengths tested. Similar trends were observed for salinities of 0.1 and 0.3 M.

It was found that the D value in cases when SDBS is used is 0.73 to 0.829, while in cases where SDBS was not employed,

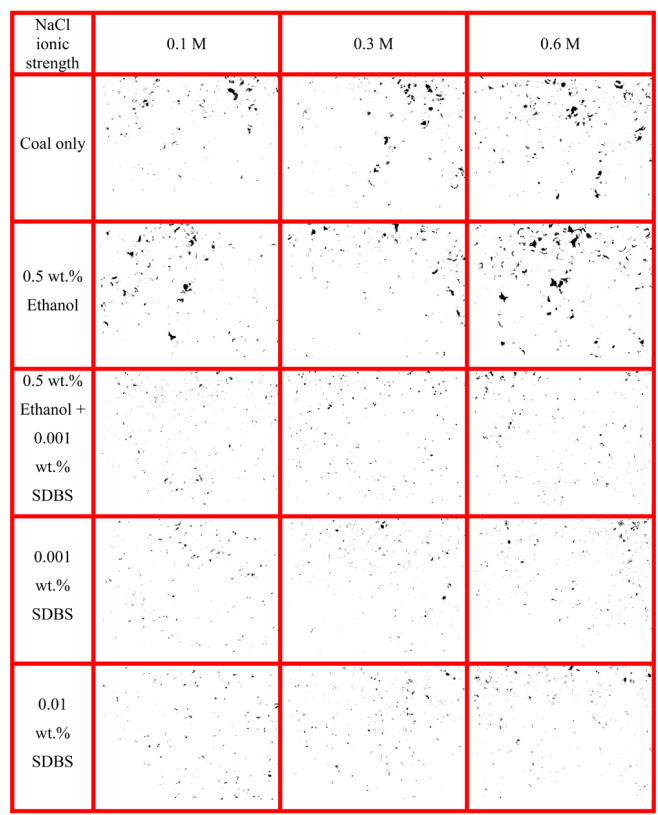

**Figure 11.** Microscopy images of the proppant pack after coal fines suspension has passed through. The pictures have a visibility window dimension of 792 × 594 μm.

the value of D remained between 0.962 and 1.122, as can be seen in Figure 12. This shows that SDBS-added suspensions demonstrated lesser aggregation than non-SDBS suspensions (S-1, S-2, S-6, S-7, S-11, and S-12), as can be seen in Figure 12. The fractal analysis result conclusively reaffirmed the fact that SDBS is an effective dispersant for coal water suspension.

These results of microscopy imaging are consistent with colloidal stability (zeta potential) observations. The implications of these results signify the dispersive effect of coal fines in the presence of SDBS. The SDBS is an effective anionic surfactant that can be an ingredient in hydraulic fracturing fluid to enhance the conductivity by recovering coal fines through the proppant pack/coal seams.

## 4. RECOMMENDATION AND SUGGESTIONS

In the current work, the SDBS surfactant is recommended for field applications to efficiently remove undesirable coal fines for the seams/proppant pack. This research is an extension on the application of surfactants to modify the surface properties of coal fines.[23,30,32,37] However, 0.001 wt % SDBS acts as an effective dispersant where coal fines are <0.1 wt % in a saline (0.1 to 0.6 M NaCl) environment having a pH level of ∼8.5. It is suggested that further research be conducted on enhanced coal fines loading (up to 5 wt %), salinity as comparable to reservoir waters, high temperature and pressure for simulating the real coal seam conditions, micromorphology,[56] nucleation kinetics[57] of coal fines surfaces treated with surfactants, coal fines movement during gas flow (as has been conducted for shales[58,59]), and facilities required at gathering systems for its optimization[60] for continuous treatment (if desired).

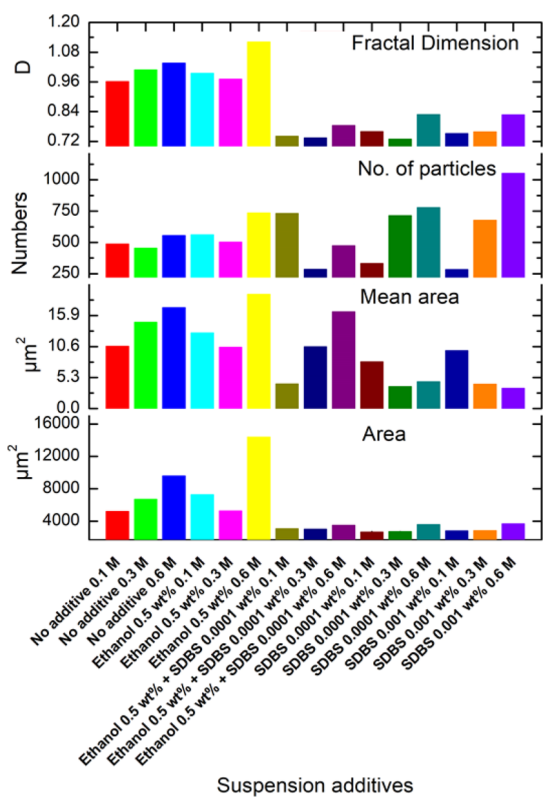

**Figure 12.** Statistical analysis of microscopy images. (a) Fractal analysis of the suspensions. (b) Number of particles in the presence of dispersants. (c) Mean area analysis in the presence of dispersants. (d) Area analysis in the presence of dispersants.

## 5. CONCLUSIONS

Coal fines detached during coal seam gas production have a dramatic impact on the hydraulic conductivity of the main advective flow path (fractures and proppant packs). We thus tested various formulations to stabilize such detached fines to minimize formation damage. From this study, the following conclusions are drawn:

(1) In order to disperse coal fines, an optimum additive concentration of 0.001 wt % SDBS (an anionic dispersant) is recommended to be added to water-based fracturing fluids. SDBS led to substantial and the most favorable enhancement of coal fines dispersion stability. Ethanol, however, did not show effective results. Note that this is related to the absolute zeta potential values of more than 50 mV when SDBS is used at salinities of 0.1 and 0.3 M. Even at standard salinity (0.6 M), the absolute zeta potential value was 34 mV, resulting in good dispersion stability. Furthermore, the zeta potential became independent of the pH when SDBS was used.

(2) The proppant pack tests showed that the formulations with no additive or ethanol as additive undergo higher fines aggregation than formulations containing SDBS. This was consistent with the effluent particle counter and FTIR measurements. Clearly, the adsorption of coal fines in the column is lower.

(3) In addition, microscopy images show that SDBS-based formulations provided excellent dispersion of coal fines in the proppant pack for all saline ranges; a further quantitative analysis of the microscopy images con-clusively depicted that SDBS-based suspensions have a significantly lower fractal dimension (<0.85) and lower area covered by coal fines (<3700 $\mu m^2$) than cases where SDBS is not added. This also demonstrates that lower aggregation of coal fines is observed when SDBS-enhanced coal suspensions are passed through the proppant packs. Eventually, it is concluded that conductivity of the proppant pack is enhanced when SDBS is used by removal of coal fines aggregation and aiding the coal fines flow through the pack with the hydraulic fracturing flowback.

Hence, higher stability of coal fines using SDBS as a dispersant can improve CSG well production by dispersing coal fines and minimizing formation damage. Thus, the coal fine migration issue in CSG can be minimized if the HF fluid is correctly formulated.

■ AUTHOR INFORMATION

**Corresponding Authors**

Faisal Ur Rahman Awan − School of Engineering, Edith Cowan University, Joondalup, Western Australia 6027, Australia; Department of Petroleum and Gas Engineering, Dawood University of Engineering and Technology, Karachi 74800, Pakistan; orcid.org/0000-0003-2394-0735; Email: f.awan@ecu.edu.au

Alireza Keshavarz − School of Engineering, Edith Cowan University, Joondalup, Western Australia 6027, Australia; orcid.org/0000-0002-8091-961X; Email: a.keshavarz@ecu.edu.au

**Authors**

Hamed Akhondzadeh − School of Engineering, Edith Cowan University, Joondalup, Western Australia 6027, Australia

Sarmad Al-Anssari − School of Engineering, Edith Cowan University, Joondalup, Western Australia 6027, Australia; Department of Chemical Engineering, University of Baghdad, Baghdad 10071, Iraq

Ahmed Al-Yaseri − School of Engineering, Edith Cowan University, Joondalup, Western Australia 6027, Australia; orcid.org/0000-0001-9094-1258

Ataollah Nosrati − School of Engineering, Edith Cowan University, Joondalup, Western Australia 6027, Australia

Muhammad Ali − School of Engineering, Edith Cowan University, Joondalup, Western Australia 6027, Australia; Department of Petroleum Engineering, Curtin University, Kensington, Western Australia 6151, Australia

Stefan Iglauer − School of Engineering, Edith Cowan University, Joondalup, Western Australia 6027, Australia

Complete contact information is available at:
https://pubs.acs.org/10.1021/acs.energyfuels.0c00045

**Notes**
The authors declare no competing financial interest.
⊥Deceased 29th of May 2019.

■ ACKNOWLEDGMENTS

This work was supported by the Higher Education Commission (HEC) Pakistan vide approval letter no. 5-1/HRD/UESTPI(Batch-V)/3371/2017/HEC and Edith Cowan University (ECU) Australia Early Career Research Grant G1003450. The authors would like to thank HEC, Pakistan and ECU for the Ph.D. grant vide ECU−HEC Joint Scholarship 2017.

## ■ NOMENCLATURE

$CS_{LSW}$ = coal suspension in low-salinity water (0.1 M NaCl brine, 0.5844 wt %)

$CS_{MSW}$ = coal suspension in medium-salinity water (0.3 M NaCl brine, 1.7532 wt %)

$CS_{HSW}$ = coal suspension in high-salinity water (0.6 M NaCl brine, 3.5064 wt %)

SDBS = sodium dodecyl benzene sulfonate (an anionic surfactant)

CMCF = chemically modified coal fines (coal fines treated with SDBS)

FTIR = Fourier transform infrared (a technique of measuring spectra)

CSG = coal seam gas (an unconventional gas reservoir)

HF = hydraulic fracturing (a process of creating fractures)

## ■ REFERENCES

(1) BP plc *BP Energy Outlook*; 2019 edition, BP group press: 2019.
(2) IEA *World Energy Outlook 2018*, I. Publications, Editor. 2018.
(3) Seidle, J. *Fundamentals of Coalbed Methane Reservoir Engineering*; PennWell Corporation: Tulsa, UNITED STATES, 2011.
(4) Li, Q.; et al. A review on hydraulic fracturing of unconventional reservoir. *Petroleum* **2015**, *1*, 8−15.
(5) Chang, Z.; et al. *Propagation Law of Hydraulic Fracture in Coal Seam Based on Element Coal Cleats Model*, in *52nd U.S. Rock Mechanics/Geomechanics Symposium*; American Rock Mechanics Association: Seattle, Washington, 2018. p. 7.
(6) Shen, J.; et al. Relative permeabilities of gas and water for different rank coals. *Int. J. Coal Geol.* **2011**, *86*, 266−275.
(7) Espinoza, D. N.; et al. Desorption-induced shear failure of coal bed seams during gas depletion. *Int. J. Coal Geol.* **2015**, *137*, 142−151.
(8) Han, G.; et al. An experimental study of coal-fines migration in Coalbed-methane production wells. *J. Nat. Gas Sci. Eng* **2015**, *26*, 1542−1548.
(9) Palmer, I. D.; Moschovidis, Z. A.; Cameron, J. R. *Coal Failure and Consequences for Coalbed Methane Wells*, in *SPE Annual Technical Conference and Exhibition*; Society of Petroleum Engineers: Dallas, Texas, 2005, 11.
(10) Akhondzadeh, H.; et al. Pore-scale analysis of coal cleat network evolution through liquid nitrogen treatment: A Micro-Computed Tomography investigation. *Int. J. Coal Geol.* **2020**, 103370.
(11) Huang, B.; et al. Hydraulic fracturing after water pressure control blasting for increased fracturing. *Int. J. Rock Mech. Min. Sci.* **2011**, *48*, 976−983.
(12) Thakur, P. Chapter 1 - Global Reserves of Coal Bed Methane and Prominent Coal Basins, in *Advanced Reservoir and Production Engineering for Coal Bed Methane*, P., Thakur, Editor, Gulf Professional Publishing: 2017, p. 1−15.
(13) Guo, Z.; Hussain, F.; Cinar, Y. Physical and analytical modelling of permeability damage in bituminous coal caused by fines migration during water production. *J. Nat. Gas Sci. Eng.* **2016**, *35*, 331−346.
(14) Akhondzadeh, H.; et al. Investigating the relative impact of key reservoir parameters on performance of coalbed methane reservoirs by an efficient statistical approach. *J. Nat. Gas Sci Eng.* **2018**, *53*, 416−428.
(15) Keshavarz, A.; et al. Enhancement of CBM well fracturing through stimulation of cleat permeability by ultra-fine particle injection. *The APPEA Journal* **2014**, *54*, 155−166.
(16) Keshavarz, A.; et al. Enhanced Gas Recovery Techniques From Coalbed Methane Reservoirs, in *Fundamentals of Enhanced Oil and Gas Recovery from Conventional and Unconventional Reservoirs*; Elsevier, 2018. p. 233−268.
(17) Keshavarz, A.; et al. Stimulation of coal seam permeability by micro-sized graded proppant placement using selective fluid properties. *Fuel* **2015**, *144*, 228−236.
(18) Tao, S.; et al. The influence of flow velocity on coal fines output and coal permeability in the Fukang Block, southern Junggar Basin. *China. Scientific Reports* **2017**, *7*, 14124.
(19) Bai, T.; et al. Experimental investigation on the impact of coal fines generation and migration on coal permeability. *Journal of Petroleum Science and Engineering* **2017**, *159*, 257−266.
(20) Wei, Y.; et al. Characteristics of Pulverized Coal during Coalbed Methane Drainage in Hancheng Block, Shaanxi Province China. *Energy Explor. Exploit.* **2013**, *31*, 745−757.
(21) Huang, F.; et al. Massive fines detachment induced by moving gas-water interfaces during early stage two-phase flow in coalbed methane reservoirs. *Fuel* **2018**, *222*, 193−206.
(22) Marcinew, R. P.; Hinkel, J. J. Coal Fines-Origin, Effects And Methods To Control Associated Damage, in *Annual Technical Meeting*; Petroleum Society of Canada: Calgary, Alberta, 1990.
(23) Zou, Y. S.; Zhang, S. C.; Zhang, J. Experimental Method to Simulate Coal Fines Migration and Coal Fines Aggregation Prevention in the Hydraulic Fracture. *Transp. Porous Media* **2014**, *101*, 17−34.
(24) Huang, T.; et al. Controlling Coal Fines in Coal Bed Operations, in *United States Patent Application Publication*, U.S.P.A. Publication, Editor. 2010, Baker Hughes Incoporated: Houston, TX (US): United States. p. 10.
(25) Shi, Q.; et al. An experimental study of the agglomeration of coal fines in suspensions: Inspiration for controlling fines in coal reservoirs. *Fuel* **2018**, *211*, 110−120.
(26) Keshavarz, A.; et al. Laboratory-based mathematical modelling of graded proppant injection in CBM reservoirs. *Int. J. Coal Geol.* **2014**, *136*, 1−16.
(27) Keshavarz, A.; et al. Stimulation of Unconventional Naturally Fractured Reservoirs by Graded Proppant Injection: Experimental Study and Mathematical Model, in *SPE/EAGE European Unconventional Resources Conference and Exhibition*; Society of Petroleum Engineers: Vienna, Austria, 2014. p. 12.
(28) Magill, D. P.; et al. Controlling Coal-Fines Production in Massively Cavitated Openhole Coalbed-Methane Wells, in *SPE Asia Pacific Oil and Gas Conference and Exhibition*; Society of Petroleum Engineers: Brisbane, Queensland, Australia, 2010.
(29) Kumar, A.; et al. Pre-fracture Treatment of Coal Seams for Fracture Conductivity Enhancement in Hydro Fracturing of CBM Wells and Coal Fines Mitigation in Multilateral CBM Wells through Wettability Alteration of Coal Fines - A Laboratory Study, in *SPE Oil and Gas India Conference and Exhibition*; Society of Petroleum Engineers: Mumbai, India, 2012. p. 9.
(30) Pan, L.-H.; et al. An experimental study on screening of dispersants for the coalbed methane stimulation. *Int. J. Oil Gas Coal Technol.* **2015**, *9*, 437−454.
(31) Ferrer, I.; Thurman, E. M. Chemical constituents and analytical approaches for hydraulic fracturing waters. *Trends Environ. Anal. Chem.* **2015**, *5*, 18−25.
(32) Chen, Y.; Xu, G.; Albijanic, B. Evaluation of SDBS surfactant on coal wetting performance with static methods: Preliminary laboratory tests. *Energy Sources, Part A* **2017**, *39*, 2140−2150.
(33) Chen, Y.; et al. Characterization of coal particles wettability in surfactant solution by using four laboratory static tests. *Colloids Surf., A* **2019**, *567*, 304−312.
(34) Cai, J.; et al. Recovery of methane from coal-bed methane gas mixture via hydrate-based methane separation method by adding anionic surfactants. *Energy Sources, Part A* **2018**, *40*, 1019−1026.
(35) Guo, M.; et al. Coal derived porous carbon fibers with tunable internal channels for flexible electrodes and organic matter absorption. *J. Mater. Chem. A* **2015**, *3*, 21178−21184.
(36) Zhao, Y.; et al. Adsorption mechanism of sodium dodecyl benzene sulfonate on carbon blacks by adsorption isotherm and zeta potential determinations. *Environ. Technol.* **2013**, *34*, 201−207.

(37) Awan, F. U. R.; et al. Optimizing the Dispersion of Coal Fines Using Sodium Dodecyl Benzene Sulfonate, in *SPE/AAPG/SEG Asia Pacific Unconventional Resources Technology Conference.*, Unconventional Resources Technology Conference: Brisbane, Australia, 2019. p. 9.

(38) Patel, A.; et al. Screening of Nanoparticles to Control Clay Swelling in Coal Bed Methane Wells, in *International Petroleum Technology Conference.*, International Petroleum Technology Conference: Bangkok, Thailand, 2016.

(39) Moore, T. A. Coalbed methane: A review. *Int. J. Coal Geol.* **2012**, *101*, 36−81.

(40) Wang, X.-J.; Zhu, D.-S.; Yang, S. Investigation of pH and SDBS on enhancement of thermal conductivity in nanofluids. *Chem. Phys. Lett.* **2009**, *470*, 107−111.

(41) Lyklema, J. Surface Chemistry of Colloids in Connection with Stability, in *The Scientific Basis of Flocculation*, K.J., Ives, Editor., Springer Netherlands: Dordrecht, 1978. p. 3−36.

(42) Buckton, G. Surface Characterization: Understanding Sources of Variability in the Production and Use of Pharmaceuticals*. *J. Pharm. Pharmacol.* **1995**, *47*, 265−275.

(43) Al-Anssari, S.; et al. Impact of nanoparticles on the CO2-brine interfacial tension at high pressure and temperature. *J. Colloid Interface Sci.* **2018**, *532*, 136−142.

(44) Pate, K.; Safier, P. 12 - Chemical metrology methods for CMP quality, in *Advances in Chemical Mechanical Planarization (CMP)*, S., Babu, Editor, Woodhead Publishing, 2016. p. 299−325.

(45) Joseph, E.; Singhvi, G. Chapter 4 - Multifunctional nanocrystals for cancer therapy: a potential nanocarrier, in *Nanomaterials for Drug Delivery and Therapy*, Grumezescu, A. M., Editor., William Andrew Publishing, 2019. p. 91−116.

(46) Carr, J. R.; Norris, G. M.; Newcomb, D. E. Characterization of aggregate shape using fractal dimension. *Transp. Res. Rec.* **1990**, *1278*.

(47) Bossler, F.; et al. Fractal approaches to characterize the structure of capillary suspensions using rheology and confocal microscopy. *J. Rheol.* **2018**, *62*, 183−196.

(48) Jha, N. K.; et al. Pore scale investigation of low salinity surfactant nanofluid injection into oil saturated sandstone via X-ray micro-tomography. *J. Colloid Interface Sci.* **2020**, *562*, 370−380.

(49) Jha, N. K.; et al. Wettability Alteration of Quartz Surface by Low-Salinity Surfactant Nanofluids at High-Pressure and High-Temperature Conditions. *Energy Fuels* **2019**, *33*, 7062−7068.

(50) Jorge Ricardo, P. Effect of the properties of coal surface and flocculant type on the flocculation of fine coal, in *Department of Mining and Mineral Process Engineering*, The University of British Columbia, 1996.

(51) Guo, Z.; Hussain, F.; Cinar, Y. Permeability variation associated with fines production from anthracite coal during water injection. *Int. J. Coal Geol.* **2015**, *147-148*, 46−57.

(52) Singh, B. P. The role of surfactant adsorption in the improved dewatering of fine coal. *Fuel* **1999**, *78*, 501−506.

(53) Arab, D.; et al. Remediation of colloid-facilitated contaminant transport in saturated porous media treated by nanoparticles. *Int. J. Environ. Sci. Technol.* **2014**, *11*, 207−216.

(54) Development, I.E.S.C.o.C.S.G.a.L.C.M *Coal seam gas extraction and co-produced water*, I.E.S.C.o.C.S.G.a.L.C.M. Development, Editor. 2014.

(55) Zhao, X.; et al. Characteristics and generation mechanisms of coal fines in coalbed methane wells in the southern Qinshui Basin, China. *J. Nat. Gas Sci. Eng.* **2016**, *34*, 849−863.

(56) Wang, Z.; et al. Formation and rupture mechanisms of viscoelastic interfacial films in polymer-stabilized emulsions. *J. Dispersion Sci. Technol.* **2019**, *40*, 612−626.

(57) Wang, Z.; et al. Investigation on gelation nucleation kinetics of waxy crude oil emulsions by their thermal behavior. *J. Nat. Gas Sci. Eng.* **2019**, *181*, 106230.

(58) Zeng, F.; et al. Gas Mass Transport Model for Microfractures Considering the Dynamic Variation of Width in Shale Reservoirs. *SPE Reservoir Eval. Eng.* **2019**, *22*, 1265−1281.

(59) Memon, K. R.; et al. Influence of Cryogenic Liquid Nitrogen on Petro-Physical Characteristics of Mancos Shale: An Experimental Investigation. *Energy Fuels* **2020**, *34*, 2160−2168.

(60) Liu, Y.; et al. Layout optimization of large-scale oil−gas gathering system based on combined optimization strategy. *Neurocomputing* **2019**, *332*, 159−183.